\documentclass[english,aps,prl,twocolumn]{revtex4}

\usepackage{amsmath}
\usepackage{amssymb}
\usepackage{color}
\usepackage{natbib}

\begin{document}


\title{Ground State Degeneracy of Interacting Spinless Fermions}
\author{Zhong-Chao Wei$^1$, Xing-Jie Han$^1$, Zhi-Yuan Xie$^1$, Tao Xiang$%
^{1,2}$}
\email{txiang@iphy.ac.cn}
\affiliation{$^1$Institute of Physics, Chinese Academy of Sciences, P.O. Box 603, Beijing
100190, China}
\affiliation{$^2$Collaborative Innovation Center of Quantum Matter, Beijing 100190, China}

\begin{abstract}
  We propose an eigen-operator scheme to study the lattice model of interacting spinless fermions at half filling and show that this model possesses a hidden form of reflection positivity in its Majorana fermion representation. Based on this observation, we prove rigourously that the ground state of this model is either unique or doubly degenerate if the lattice size $N$ is even, and is always doubly degenerate if $N$ is odd. This proof holds in all dimensions with arbitrary lattice structures.
\end{abstract}

\maketitle

As a fundamental model of strongly correlated systems, interacting spinless fermions have been studied for many decades.
Interest in the investigation on this system has revived recently in connection with the study of the quantum critical behavior of Dirac fermions\cite{2014HoffmanPRB,2014WangNJP,2014YaoHong,2014YaoHong2} and Majorana fermions in topological superconductors \cite{2009Kitaev, 2014Nadj-Perge}.
The spinless fermion model provides a minimal realization of lattice Dirac fermions coupled to the Ising order parameter, whose quantum critical behavior at low energy can be described by the effective Gross-Neveu-Yukawa theory \cite{1974GrossPRD,2006HerbutPRL,2009HerbutPRB1,2009HerbutPRB2}.
This model is used to mimic the charge dynamics of the extended Hubbard model in the strong coupling limit\cite{1990SchulzPRL,2000DiasPRB,2004KivelsonPRB}, and to solve some puzzling phenomenona related to the electronic structure of cuprates and organic superconductors, for example the phase separation, the incommensurate charge-density wave\cite{1993UhrigPRL}, the stripe order\cite{2001HenleyPRB,2003ZhangPRB} and the nematic phase\cite{2004KivelsonPRB}.
It is also introduced to describe and the spin-polarized electrons in ferromagnetic materials, for example in the study of the Verwey transition of magnetite (Fe$_3$O$_4$)\cite{1971CullenPRL} and some  Mn-based Heusler alloys\cite{1983deGrootPRL}, and the many-body localization effect\cite{1998SchmitteckertPRL,2015LevPRL}.
On frustrated lattices this system shows exotic phenomenon, such as the charge fractionalization\cite{2006PollmanPRB} and the time-reversal symmetry breaking\cite{2013TielemanPRL}.
In one dimension, the spinless fermion model with nearest neighbor interactions is equivalent to the anisotropic Heisenberg spin chain and can be rigorously solved using the Bethe Ansatz.

The model of interacting spinless fermions considered in this paper is defined on a bipartite lattice,
\begin{eqnarray}
H & =& H_K + H_V ,\label{Eq:ModelOrigin} \\
H_K & = & \sum_{ij}2t_{ij}(c_{i}^{\dagger }c_{j}+h.c.), \nonumber \\
H_V & = & \sum_{ij} V_{ij}(2n_{i}-1)(2n_{j}-1) ,
\end{eqnarray}
where $n_i=c^\dagger_i c_i$, $\{t_{ij}\}$ are the real nearest-neighbour hopping coefficient for fermions and $\{V_{ij}\}$ stand for interactions between different sites. $V_{ij}\geq 0$ if $i$ and $j$ belong to different sublattices, and $V_{ij}\leq 0$ if $i$ and $j$ belong to the same sublattice.
If $V_{ij}\neq 0$, we say that there is a bond between $i$ and $j$. We assume the lattice to be \emph{connected} in which there is a connected path of bonds between every pair of sites.
This Hamiltonian is defined at the half filling. It possesses a particle-hole symmetry.

The above Hamiltonian looks simple. But it is not exactly soluble except in one dimension.
In this work, we present a rigorous proof for two theorems on the ground state degeneracy of this model.
We will show that the ground state is doubly degenerate when the lattice size $N$ is odd and at most doubly degenerate when $N$ is even.
Our proof is based on the observation of a hidden symmetry of reflection positivity in the Hamiltonian.
It starts by introducing the Majorana fermion representation for the fermion operators to convert Eq. (\ref{Eq:ModelOrigin}) into an interacting model of Majorana fermions.
In this representation, each fermion is represented by two Majorana fermions.
The Hamiltonian is reflection symmetric under the exchange of these two kinds of Majorana fermions.
Under the framework of eigen-operators, we show that this symmetry puts strong constraint on the sign structure of the ground state wavefunction and can be used to determined the number of degeneracy for the ground states.

The method of reflection positivity was first developed in quantum field theory.
It has found many applications in both classical and quantum statistical physics\cite{1976FrohlichCMP,1978DysonJSP,1978FrohlichCMP}.
In 1989, Lieb proved that the ground state of the Hubbard model is unique by utilizing a Perron-Frobenius-type argument based on the spin-reflection positivity of the system\cite{1989LiebPRL}. His work extended the proof for the uniqueness of the ground state from an antiferromagnetic Heisenberg model\cite{1962Lieb} to an interacting fermion system and led to a more general application of reflection positivity in the flux phase problem\cite{1994LiebPRL}, frustrated Heisenberg antiferromagnets\cite{1999LiebSchuppPRL}, and other quantum lattice models\cite{2004TianJSP}.
The spin-reflection positivity is intimately connected with the sign rule of the ground-state wave function, which can be used to understand the minus sign problem caused by the fermion characteristic of electrons in quantum Monte Carlo simulations\cite{2004TianJSP}.
The reflection positivity for Majorana fermions was first studied in Refs. \cite{2015Jaffe} and applied to a Majorana fermion model with topological order in Refs. \cite{2014JaffeEPL,2015TerhalPRL}.

\emph{Majorana representation:}
A spinless fermion can be decomposed as two Majorana fermions.
To define $c_{i} = \left( \gamma _{i}^{\left( 1\right) }+i\gamma _{i}^{\left( 2\right) } \right)/2$ on one sublattice, and $c_{i}=\left( \gamma _{i}^{\left( 2\right) }+i\gamma _{i}^{\left( 1\right) } \right)/2$ on the other sublattice, we can rewrite Eq. (\ref{Eq:ModelOrigin}) as a two-component Majorana fermion model
\begin{eqnarray}
H_{K} &=&\sum_{ij}t_{ij}\left( i\gamma _{i}^{\left( 1\right) }\gamma
_{j}^{\left( 1\right) }-i\gamma _{i}^{\left( 2\right) }\gamma _{j}^{\left(
2\right) }\right) ,  \nonumber \\
H_{V} &=&-\sum_{ij}V_{ij}\left( i\gamma _{i}^{\left( 1\right) }\gamma
_{j}^{\left( 1\right) }\right) \left( -i\gamma _{i}^{\left( 2\right) }\gamma
_{j}^{\left( 2\right) }\right) ,  \nonumber
\end{eqnarray}
where $\gamma _{i}^{\left( \sigma \right) }$ ($\sigma =1,2$) are Majorana fermion operators at site $i$. We have made the substitution $V_{ij}\rightarrow -V_{ij}$ for the coupling constants between different sublattices, thus all $V_{ij}\geq 0$.

Majorana fermion operators are self-conjugate operators. To discuss the degeneracy of ground states, it is more convenient to use the formulation of eigen-operators rather than that of eigen-vectors.
This is because in the eigenvector formulation, the total degrees of freedom for Majorana fermions have to be doubled for odd $N$.
This doubling of degrees of freedom can be avoid in the eigen-operator representation since at each lattice site $i$, $\left( 1,\gamma _{i}^{(1)},\gamma_{i}^{(2)}, i\gamma _{i}^{(1)}\gamma _{i}^{(2)}\right) $ form a complete orthonormal basis set for any operators defined at this site.
The direct product of these operators over all lattice sites form a complete basis set
for all operators defined on the lattice.

Using the eigen-vectors of the Hamiltonian, it is easy to construct eigen-operators $O$ in each sector distinguished by symmetry, which satisfy the following operator eigen-equation
\begin{equation}
HO=OH=EO,
\end{equation}%
where $E$ is the eigen-energy. In the subspace spanned by $n$ degenerate
eigen-vectors of $H$ with eigenvalue $E$, say $\Omega _{E}=\left\{
\left\vert \Psi _{a}\right\rangle ,a=1,\cdots ,n\right\}$, $O$ can be written as a superposition of $n^{2}$ linear independent
operators $\left\{\left\vert \Psi _{a}\right\rangle \left\langle \Psi
_{b}\right\vert\right\}$, where both $\left\vert \Psi _{a}\right\rangle $ and $%
\left\vert \Psi _{b}\right\rangle $ belong to $\Omega _{E}$. Thus for each energy level with distinct good quantum numbers, the degeneracy of eigenvectors is equal to
the square root of the degeneracy of the corresponding eigen-operators.
Moreover, each eigen-operator can always be symmetrized(or antisymmetrized) to be a Hermitian
operator $O^{\dagger }=O$.

The Hamiltonian preserves the parity (even or odd) of the number of $\gamma
^{(\sigma )}$ fermions. Thus the eigen-operators can be diagonalized into four blocks
according to the parities of $\gamma ^{(1)}$ and $\gamma ^{(2)}$ fermions. Operators in each parity sector can be expanded
using the basis operators defined by%
\begin{eqnarray}
\Gamma _{\alpha }^{(1)} &=&i^{[m/2]}\gamma _{i_{1}}^{(1)}\cdots \gamma
_{i_{m}}^{(1)}, \\
\Gamma _{\alpha }^{(2)} &=&\left( -i\right) ^{[m/2]}\gamma
_{i_{1}}^{(2)}\cdots \gamma _{i_{m}}^{(2)},
\end{eqnarray}%
where $\alpha =\left( i_{1},\cdots ,i_{m}\right) $ denotes the configuration
of Majorana fermions. $[m/2]=m/2$ or $(m-1)/2$ if $m$ is even or odd. We say $%
\Gamma _{\alpha }^{\left( \sigma \right) }$ is even or odd, if $%
\alpha $ contains an even or odd number of Majorana fermion. The operators
such defined are Hermitian and form a complete orthonormal basis set of
operators.
Moreover, the Hamiltonian commutes with the following two parity operators
\begin{eqnarray}
\Delta ^{(1)} &=&i^{[N/2]}\gamma _{1}^{\left( 1\right) }\cdots \gamma
_{N}^{\left( 1\right) }, \\
\Delta ^{(2)} &=&(-i)^{[N/2]}\gamma _{1}^{\left( 2\right) }\cdots \gamma
_{N}^{\left( 2\right) }.
\end{eqnarray}
$\Delta ^{\left( 1\right) }$ commutes or anticommutes with $\Delta ^{\left(
2\right) }$ for even or odd $N$.

An eigen-operator can be expanded using the basis operators as
\begin{equation}
O=\sum_{\alpha \beta }\Psi _{\alpha \beta }\Gamma _{\alpha }^{(1)}\Gamma
_{\beta }^{(2)}.  \label{Eq:Operator1}
\end{equation}%
In the even-even sector, $\left\{\Psi _{\alpha \beta }\right\}$ is a Hermitian matrix satisfying the following eigen-equation
\begin{equation}
K\Psi +\Psi K-\sum_{ij}V_{ij}L_{ij}\Psi L_{ij}=E\Psi ,  \label{Eq:EigenEq}
\end{equation}%
where $K$ and $L_{ij}$ are Hermitian matrices defined by
\begin{eqnarray}
L_{ij,\alpha ^{\prime }\alpha } &=&\mathrm{Tr}\left[ \Gamma _{\alpha ^{\prime
}}^{(1)}i\gamma _{i}^{\left( 1\right) }\gamma _{j}^{\left( 1\right) }\Gamma
_{\alpha }^{(1)}\right] , \\
K_{\alpha ^{\prime }\alpha } &=&\sum_{ij}t_{ij}L_{ij,\alpha ^{\prime }\alpha
}.
\end{eqnarray}%
The real part of $\Psi$ represents the symmetric part of the operator under the reflection operation, while the imaginary part stands for the antisymmetric part.
The eigen-energy is given by%
\begin{equation}
E\left( \Psi \right) =2\mathrm{Tr}\left( K\Psi^2 \right) -\sum_{ij}V_{ij}\mathrm{Tr}\left(
\Psi L_{ij}\Psi L_{ij}\right) ,  \label{Eq:Energy}
\end{equation}%
if the eigen-operator $\Psi $ is normalized, $\mathrm{Tr}\left( \Psi ^{2}\right) =1$.

Below we discuss about the degeneracy of the ground states.
For clarity, we consider the cases of odd and even lattice size separately.

{\it Odd lattice size system:}
In the case $N$ is odd, the parity operator $\Delta ^{\left( \sigma \right) }$ contains an odd number of Majorana fermions.
It defines a transformation between two sectors with opposite parities for the  $\gamma
^{(\sigma )}$ fermions.
For any given eigen-operator in the even-even sector $O$, $\Delta ^{\left(
1\right) }O$ generates an operator in the odd-even sector. Similarly,
operators in the even-odd and odd-odd sectors can be generated by applying $%
\Delta ^{\left( 2\right) }$ and $\Delta ^{\left( 1\right) }\Delta ^{\left(
2\right) }$ to $O$, respectively. Since both $\Delta ^{\left( 1\right) }$
and $\Delta ^{\left( 2\right) }$ commute with the Hamiltonian, it is simple
to show that all these four operators are degenerate eigen-operators of $H$. Thus if the ground state eigen-operator is $n$-fold
degenerate in the even-even parity sector, then the total degeneracy of the
eigen-operators is $4n$ and the ground state degeneracy is $2\sqrt{n}$. This
leads to the following theorem:

{\bf Theorem I:} For the model defined by Eq. (\ref{Eq:ModelOrigin}), the
ground state is doubly degenerate if $N$ is odd.

To proof this theorem, we only need to show that the ground state is non-degenerate,
i.e., $n=1$, in the even-even parity sector. This can be done in three steps: (1) To show that among all degenerate ground states in the even-even parity sector there always exists an eigenfunction $\Psi$ which is semi-positive definite. (2) To show that all the basis states in the even-even parity sector are connected if the whole lattice is connected by the hopping or interacting terms. This means that $\Psi$ must be a full-rank matrix. (3) To show that the ground state is unique in this parity sector.

\emph{Proof: }
Let us assume $\Psi $ to be a Hermitian matrix that represents an eigen-operator of the
Hamiltonian, satisfying Eq. (\ref{Eq:EigenEq}). By diagonalizing it using an
unitary matrix $U$, $\Psi =$ $U\Lambda U{}^{\dagger }$, we can define a
trial wave function $|\Psi |=U|\Lambda |U{}^{\dagger }$ by setting the
diagonal matrix $\Lambda $ to its absolute value. $\left\vert \Psi
\right\vert $ is semi-positive definite as defined. From Eq. (\ref{Eq:Energy}), it is
simple to show that
\begin{equation}
E(\left\vert \Psi \right\vert )\leq E(\Psi ),
\end{equation}%
if not all $ V_{ij}  $ are vanished. Thus if $\Psi $ is a ground state eigen-operator, so is $\left\vert \Psi \right\vert $.

With $\Psi $ given then, now let us consider the sime-positive matrix $%
R=\left\vert \Psi \right\vert -\Psi $ which is also a multiple of a ground
state and satisfies (\ref{Eq:EigenEq}). We denote $\Omega $ as the ensemble
of all vectors $v$ that satisfy $Rv=0$. For a given $v$ in $\Omega $, by
acting $v^{\dagger }$ and $v$ to both sides of the eigen-equations, it can
be shown that $RL_{ij}v=RL_{ij}^{T}v=0$ for all connected bonds, hence $%
L_{ij}v$ and $L_{ij}^{T}v$ are also in $\Omega $. Note that $v$ corresponds
to an operator $r(v)=\sum_{\alpha =even}v_{\alpha }\Gamma _{\alpha }^{\left(
\sigma \right) }$ in the even-parity sector. The actions of $L_{ij}v$ and $%
L_{ij}^{T}v$ are equivalent to multiplying $r(v)$ by $\Gamma _{ij}^{\left(
\sigma \right) }=i\gamma _{i}^{\left( \sigma \right) }\gamma _{j}^{\left(
\sigma \right) }$ from its right and left hand side, respectively. As the
lattice is connected, all basis operators $\Gamma _{\alpha }^{\left( \sigma
\right) }$ in the even-parity sector can be generated from the product of a
set of $\Gamma _{ij}^{\left( \sigma \right) }$ on connected bonds. $\Gamma
_{ij}^{\left( \sigma \right) }$ itself is also a basis operator.
If $r(v)$ is non-zero, one can always convert it into the unity operator by
applying successively the basis operators to it, from either left or right
hand side. This is because $\left\{ \Gamma _{ij}^{\left( \sigma \right)
},\Gamma _{\alpha }^{\left( \sigma \right) }\right\} =0$ if $i\in \alpha $
and $j\notin \alpha $. Thus for any given $\alpha $ with $v_{\alpha }\neq 0$
and ($i\in \alpha $, $j\notin \alpha $), one can remove this term from $r(v)$
by the transformation $r(v)\rightarrow \Gamma _{ij}^{\left( \sigma \right)
}r(v)\Gamma _{ij}^{\left( \sigma \right) }+r(v)$. Repeating this step by step, we
can reduce $r(v)$ to a single basis operator with a constant coefficient,
which in turn can be further reduced to an unity operator by multiplying its
inverse. Thus starting from any given $v\neq 0$ in $\Omega $, one can
generate all other vectors in $\Omega $. This implies that $\Psi =\pm
\left\vert \Psi \right\vert $ and $\Psi $ is positive definite.

If there are two normalized ground states, $\Psi _{1}$ and $\Psi _{2}$ with $%
\Psi _{1}\not=\pm \Psi _{2}$, then for any real constant $p$, $\Psi _{p}=\Psi
_{1}+p\Psi _{2}$ is also a ground state. It is simple to verify that there
exists a $p$ for which $\Psi _{p}$ is neither positive nor negative
semidefinite. This contradicts the assertion that $\Psi _{p}=\pm \left\vert
\Psi _{p}\right\vert $. Thus the ground state eigen-operator is unique in the even-even
parity sector. \emph{Q.E.D}.

When $N$ is odd, the ground state can be set as the eigenstate of $\Delta
=\Delta ^{\left( 1\right) }\Delta ^{\left( 2\right) }$ with the eigenvalues,
$\pm 1$. But the eigen-operators of $\Delta $ do not preserve the parity of
Majorana fermions.

{\it Even lattice size system:}
In the case $N$ is even, both $\Delta ^{\left( 1\right) }$ and $\Delta
^{\left( 2\right) }$ are good quantum numbers and the eigen-states can be
classified by their eigenvalues. As $\Delta ^{(\sigma )}$ ($\sigma =1$, $2$) preserves the parity of $\gamma^{(\sigma )}$ fermions and the square of its eigenvalue is always equal to 1, the eigen-operator in each subspace should commute with $\Delta ^{(\sigma )}$. The basis operator $\Gamma _{\alpha }^{\left( \sigma \right) }$ commutes or anticommtes with $\Delta ^{(\sigma )}$, when it is even or odd. The operators in the even-odd, odd-even, or odd-odd parity sectors anticommute with at least one of the operators in $\Delta ^{(1)}$ and $\Delta ^{(2)}$, and thus cannot be the eigen-operators of the Hamiltonian. Thus we can discuss the ground states using only the even-even parity sector.

In the even-even parity sector, the eigen-operators can be
block-diagonalized according to the eigenvalues of $\Delta ^{(1)}$ and $%
\Delta ^{(2)}$ into four blocks. The basis operators are now defined by $%
\Gamma _{\alpha }^{(\sigma ,\pm )}=c_{\alpha ,\pm }\left[ 1\pm \Delta
^{(\sigma )}\right] \Gamma _{\alpha }^{\left( \sigma \right) }$ and the
eigen-equation becomes%
\begin{equation}
K^{\mu }\Psi _{\mu \nu }+\Psi _{\mu \nu }K^{\nu }-\sum_{ij}V_{ij}
L_{ij}^{\mu }\Psi _{\mu \nu }L_{ij}^{ \nu }=E\Psi _{\mu \nu },
\label{Eq:EigenEq2}
\end{equation}%
where $c_{\alpha ,\pm }$ is a normalization constant, $\Psi _{\mu \nu }$
with $\mu =\pm $ and $\nu =\pm $ represent the eigen-operators, and
\begin{eqnarray}
L_{ij,\alpha ^{\prime }\alpha }^{\mu } &=&\mathrm{Tr}\left[ \Gamma _{\alpha ^{\prime
}}^{(1,\mu )}i\gamma _{i}^{\left( 1\right) }\gamma _{j}^{\left( 1\right)
}\Gamma _{\alpha }^{(1,\mu )}\right] , \\
K_{\alpha ^{\prime }\alpha }^{\mu } &=&\sum_{ij}t_{ij}L_{ij,\alpha ^{\prime
}\alpha }^{\mu}.
\end{eqnarray}

\textbf{Theorem II: } For the model defined by Eq. (\ref{Eq:ModelOrigin}), there are at most two linear independent ground states labeled by $\Delta
^{(1)}=\Delta ^{(2)}=\pm 1$ when $N$ is even.

To prove this theorem, again we only need to consider the even-even parity sector. But the ground state can now fall into any of the four blocks characterized by the eigenvalues of $\Delta ^{(1)}$ and $\Delta ^{(2)}$.
Following the same steps presented in the proof of Theorem I, it is straightforward to show that the lowest energy state is unique in each of the $\Delta ^{(1)}=\Delta ^{(2)}= \pm 1$ blocks. If we can further show that the ground state does not lie in the blocks with $\Delta ^{(1)} \not= \Delta ^{(2)}$, the theorem is then proven. Below we give such a proof.

\emph{Proof: }
From the symmetry between the two kinds of fermions, it is straightforward
to show that the lowest energy state of $\Delta ^{(1)}=-\Delta ^{(2)}=1$ is degenerate with that of $\Delta ^{(2)}=-\Delta ^{(1)}=1$. Moreover, if $\Psi _{+-}=W$ is the lowest energy eigen-functon in the $\Delta ^{(1)}=-\Delta ^{(2)}=1$ block, then the corresponding eigen-functoin in the $\Delta ^{(1)}=-\Delta ^{(2)}=1$ block is given by $\Psi_{-+}=W^{\dagger}$. By taking the singular value decomposition for $W=U\Lambda V^{\dagger }$, we can introduce two trial wave functions $\Psi _{++}=U\Lambda U^{\dagger }$ and $\Psi _{--}=V\Lambda V^{\dagger }$ defined in the $\Delta ^{(1)}=\Delta^{(2)}=\pm 1$ blocks, respectively. Here $U$ and $V$ are two unitary matrices
and $\Lambda $ is a positive semidefinite diagonal matrix.

Using Eq. (\ref{Eq:EigenEq2}), it can be shown that
\begin{equation}
E(\Psi _{+-})+E(\Psi _{-+})\geq E(\Psi _{++})+E(\Psi _{--}) ,
\end{equation}%
where the equality holds if and only if $\Lambda U^{\dagger }L_{ij}^{\left(
+\right) }U=\Lambda V^{\dagger }L_{ij}^{(-)}V$ on all bonds with $V_{ij}>0$.
This means that if $\Psi _{+-}$ is a ground state, so are $\Psi _{++}$ and $%
\Psi _{--}$. Since the ground state in the $\Delta ^{(1)}=\Delta ^{(2)}=1$
is unique, $\Lambda $ should be strictly positive. Thus we have
\begin{equation}
U^{\dagger }L_{ij}^{\left( +\right) }U=V^{\dagger }L_{ij}^{(-)}V
\end{equation}%
for all $V_{ij}>0$. Since $\Delta ^{(\sigma )}$ can always be constructed
from a product of $\left( i\gamma _{i}^{\sigma }\gamma _{j}^{\sigma }\right)
$ on a set of connected bonds, this equation implies that $U^\dagger U = -V\dagger V$, which is clearly not valid.
Thus the ground state exists only in the $\Delta ^{(1)}=\Delta ^{(2)}=\pm
1 $ blocks and is at most two-fold degenerate. The degeneracy occurs if and
only if these two states are degenerate. \emph{Q.E.D}.

Thus for even $N$, the ground state is either unique or doubly degenerate.
The degeneracy happens if the Hamiltonian is invariant under a symmetry
transformation $T$ but $\Delta ^{\left( \sigma \right) }$ is odd under this
transformation, i.e., $T\Delta ^{\left( \sigma \right) }T^{\dagger }=-\Delta
^{\left( \sigma \right) }$. Some examples in which the ground state is
doubly degenerate are

(1) A system defined on a $L_{1}\times L_{2}\times \ldots \times L_{n}$
lattice with translational invariance (the periodic boundary condition is
implicitly assumed) along the first dimension (not necessary to be invariant
along any other dimension), if $L_{1}$ is even and $L_{2}\times \ldots
\times L_{n}$ is odd, then the ground state is two-fold degenerate.

(2) A system of $N=4m-2$ ($m$ is a positive integer) which is invariant
under a one-to-one correspondence mapping between half of the lattice sites
with the other half ones, the ground state is two-fold degenerate. This
includes the systems with center reflection or mirror reflection symmetries.

{\it Discussion and summary:}
The above two theorems can be extended to other interacting fermion models.
For example, the Hamiltonian can be extended to include a staggered magnetic field
\begin{equation}
H_{\mathrm{stagg}} = -\sum_{i} h_{i}(2n_{i}-1),
\label{Eq:stagg}
\end{equation}
where $h_i \geq 0$ on one sublattice and $h_i \leq 0$ on the other sublattice.
It is simple to show that the full Hamiltonian still preserves the reflection positivity.
But the different parity sectors of Majorana fermions are mixed by this newly added term. In particular, the even-even sector is mixed with the odd-odd sector, and the even-odd sector is mixed with the odd-even sector. In this case, it can be shown that the ground state is always unique, no matter whether the lattice size is even or odd. (A detailed proof on this is given in the Supplementary Information). The relative phase of the wave function between the two mixed sectors is determined by the sign structure of $h_i $.

Theorem II can be also extended to apply to the Hubbard model on a bipartite lattice at half filling. To do this, we need first convert the Hubbard model into an interacting spinless fermion model by regarding the up and down spin electrons as spinless fermions but defined on different sublattices. Thus in the language of spinless fermions, the lattice size is doubled. If A and B are the two sublattices in the original lattice, then there are four sublattices in the corresponding spinless fermion system. They can be denoted as A-up, A-down, B-up and B-up sublattices, respectively. If we group A-up and B-down as one sublattice and A-down and B-up as another in the new bipartite lattice, the Hubbard model is just a spinless fermion model as defined by Eq. (\ref{Eq:ModelOrigin}). Only some of the hoping and interacting terms are absent. But this does not affect the lattice connectivity. Direct application of Theorem II to the Hubbard model indicates that the ground state is at most doubly degenerate, consistent with the theorem proven by Lieb\cite{1989LiebPRL}.

The reflection positivity of the Hamiltonian determines the sign structure of the ground state wavefunction. In Refs. \cite{2014HoffmanPRB,2014WangNJP,2014YaoHong}, it was shown that there is no minus sign problem in the quantum Monte Carlo simulations for the model defined by Eq. (\ref{Eq:ModelOrigin}) at half filling. It can be shown that the minus sign problem is inherently connected with the reflection positivity. In particular, it can be shown that this system is minus sign free using two inequalities proven by Jaffe and Janssens for the interacting fermions with reflection positivity in the Majorana representation\cite{2015Jaffe2}, consistent with the claim made in Refs. \cite{2014HoffmanPRB,2014YaoHong} A detailed discussion on this will be published separately\cite{WeiWu}.

To summarize, we prove two theorems on the ground state degeneracy for a spinless fermion model with short-range interactions. The proof is based on reflection positivity of Majorana fermions, which can also be applied to other interacting systems with non-Abelian statistics, such as  models with parafermions\cite{2015JaffeCMP}. It is also interesting to generalize this work to systems with more than two species of Majorana fermions, such as $SU(N)$ or $O(2N)$ fermion models.

We would like to thank Dr. Ziyang Meng and Dr. Lei Wang for useful discussion. This work was supported by the National Natural Science Foundation of China (Grants No. 11190024 and No. 11474331) and by the National Basic Research Program of China (Grants No.
2011CB309703).


\section*{SUPPLEMENTAL INFORMATION}
\renewcommand{\theequation}{S.\arabic{equation}}
\setcounter{equation}{0}

Here we provide a proof for the non-degeneracy of the ground state in the presence of staggered external fields.

In the presence of staggered external fields, the Hamiltonian is defined by
\begin{equation}
H = H_K + H_V + H_{\mathrm{stagg}},
\label{Eq:ModelStagg}
\end{equation}
where $H_K$ and $H_V$ are the hopping and interaction terms defined in Eq. (1). $H_{\mathrm{stagg}}$ is the interaction defined by Eq. (19). In the Majorana representation, it reads
\begin{equation}
H_{\mathrm{stagg}} = - i \sum_{i} |h_{i}| \gamma _{i}^{\left( 1\right)} \gamma _{i}^{\left( 2\right)} ,
\end{equation}
where $|h_{i}|\ge 0$ and is nonzero at least on one lattice site. The ground state of this model is unique no matter the lattice size is odd or even. Below we give a proof for this.

In the presence of the staggered fields, neither $\Delta ^{\left(1\right) }$ nor $\Delta ^{\left(2\right) }$ commutes with the Hamiltonian, but $\Delta = \Delta ^{\left(1\right) }\Delta ^{\left(2\right) }$ does. Thus the ground state should be a common eigenstate of $H$ and $\Delta$. To count the degeneracy, we only need to consider the  eigen-operators which commute with $\Delta$.

The staggered field mixes the operators in the even-even sector with those in the odd-odd sector. Similarly, the operators in the even-odd and odd-even sectors are mixed. Since the operators in the even-odd and odd-even sectors do not commute with $\Delta$, they cannot be the simultaneous eigen-operators of the Hamiltonian and $\Delta$. Thus we only need to consider the eigen-operators in the even-even and odd-odd sectors. Generally the eigen-operator of the Hamiltonian can be expressed using the basis operators as
\begin{equation}
O = \sum_{\alpha ,\beta \in \mathrm{even} }\Psi^{\left(\mathrm{ee}\right)}_{\alpha \beta }\Gamma _{\alpha }^{(1)}\Gamma
_{\beta }^{(2)} + \sum_{\alpha ,\beta \in \mathrm{odd} } i\Psi^{\left(\mathrm{oo}\right)}_{\alpha \beta }\Gamma _{\alpha }^{(1)}\Gamma
_{\beta }^{(2)},
\end{equation}
where $\Psi^{(\mathrm{ee})}$ and $\Psi^{(\mathrm{oo})}$ are the eigen-functions in the even-even and odd-odd sectors, respectively. Both $\Psi^{(\mathrm{ee})}$ and $\Psi^{(\mathrm{oo})}$ are Hermitian matrices.

The ground state is a functional of $\Psi = \left(\Psi^{(\mathrm{ee})}, \Psi^{(\mathrm{oo})} \right)$. The eigen-equation for $\Psi$ can be similarly derived as for Eq. (9). But $\Psi^{\left(\text{ee}\right)}$ and $\Psi^{(\mathrm{oo})}$ are now coupled together by the external fields.

Following the steps given in the proof of Theorem I, it is straightforward to show that the above Hamiltonian still preserves the reflection positivity in the Majorana representation and
\begin{equation}
E(|\Psi^{(\mathrm{ee})}|, | \Psi^{(\mathrm{oo})} | ) \leq E(\Psi^{(\mathrm{ee})}, \Psi^{(\mathrm{oo})}  ).
\end{equation}
Again $|\Psi^{(\mathrm{ee})}|$ is a matrix of semi-positive definite, defined from $\Psi^{(\mathrm{ee})}$ by setting all eigenvalues of $\Psi^{(\mathrm{ee} )}$ to its absolute values. $| \Psi^{(\mathrm{oo})} |$ is similarly defined.

Furthermore, by setting $R^{(\mathrm{ee})}= |\Psi^{(\mathrm{ee})} | - \Psi^{(\mathrm{ee})}$, $R^{(\mathrm{oo})}=| \Psi^{(\mathrm{oo} )} | -\Psi^{(\mathrm{oo})}$, and assuming $R^{(\mathrm{ee})}v=0$ or $R^{(\mathrm{oo} )}v=0$ for some vectors $v$, from the eigen-equation of $(\Psi^{(\mathrm{ee})}, \Psi^{(\mathrm{oo})}  )$ and the connectivity of the lattice (or the ergodicity of states) it can be shown that the set of $v$ that satisfies the two equations should either be empty or contain all basis vectors in the even-even and odd-odd subspaces. This implies $\Psi = \pm |\Psi |$. Thus $\Psi$ is positive definite up to a global phase factor and the ground state is unique.

\end{document}